\documentstyle [12pt]{article}

\input epsf
\def\ni{\noindent}
\def\N{{\cal N}}

\begin{document}

\title{\bf Studies of Diffusional Mixing in Rotating Drums via
           Computer Simulations}
 
\author{G.A. Kohring\\
        Central Institute for Applied Mathematics\\
        Research Center J\"ulich (KFA)\\
        D-52425 J\"ulich, Germany\\
        g.kohring@kfa-juelich.de}

\date{ }
 
\maketitle
 
\begin{abstract}

Particle diffusion in rotating drums is studied via computer simulations
using a full 3-D model which does not involve any arbitrary input
parameters. The diffusion coefficient for single-component systems agree
qualitatively with previous experimental results.
On the other hand, the diffusion coefficient for two-component systems
is shown to be a highly nonlinear function of the rotation velocity. It
is suggested that this
results from a competition between the diffusive motion parallel to,
and flowing motion perpendicular to the axis of rotation. \\ 

\begin{center} 
(submitted to {\em Journal de Physique} )
\end{center} 

\end{abstract}

\section{Introduction}

Rotating drums perform critical functions in many industrial processes.
They are utilized for tasks such as humidification, dehumidification,
convective heat transfer, facilitation of gas-solid catalytic reactions and,
not the least, ordinary mixing or blending. Understanding the 
mechanisms behind particle mobility in rotating drums is an important step
towards refining the efficiency of such processes. 
In the absence of mechanical agitators, momentum is imparted to the particles 
only in the plane perpendicular to the rotation axis.
Motion parallel to the axis of rotation occurs through momentum changes
induced by the collisions between particles. Since these collisions occur
more or less randomly in time, this motion is diffusive in nature.
Until now most computer simulations have
neglected the essential three-dimensional character of these systems and,
via two-dimensional simulations,
concentrated on the motion perpendicular to the rotation axis \cite{2D}. 
In the present
work the diffusive motion in the longitudinal direction is examined using a 
full three-dimensional model.

Previous experimental studies of diffusive motion in rotating cylinders have
concentrated on single-component systems, i.e., a cylinder partially filled
with a single particle type \cite{HOGG,RAO}. 
One-half of the particles were dyed a different
color in order to make them distinguishable. 
These diffusion experiments were simplified through the use of an 
ideal starting arrangement: initially, the colored particles are axially
segregated so that mixing occurs
only through the diffusive motion parallel to the rotation axis. Under 
these ideal circumstances, some aspects of the diffusive mixing can be 
studied analytically and Hogg et al. \cite{HOGG} obtained good agreement 
between the theory and experiment.

The present study focuses on diffusion in two-component systems. 
Diffusion in two-component systems is expected to exhibit
characteristics not found in single-component systems, because the
former have been shown to exhibit spontaneous material segregation
\cite{MATSEG}. In this 
form of segregation several bands parallel to the rotation axis are formed.
These bands alternate between the two material types and there is normally
an odd number of such bands.
The nature of this phenomena is not completely understood, however, it is 
clear that diffusion must play an important role.
Indeed the results to be presented here demonstrate a distinct difference 
between the diffusion in 1-component and 2-component systems.

In the following section, a three dimensional model for colliding,
visco-elastic spheres is presented and its numerical implementation is
discussed. The next section describes the experimental arrangement. Following
that, the results are presented and discussed. The paper concludes with
some speculative comments on the mechanisms behind material segregation in
rotation drums.

\section{A Computational Model for Visco-Elastic, Mesoscopic, Spherical 
Particles}

The contact-force model presented here is
based upon a visco-elastic model for the collision of two
mesoscopic, spherical particles developed over the last 100 years by Hertz
\cite{HERTZ}, 
Johnson et al. \cite{MESO}, Kuwabara and Kono \cite{KK} and Mindlin
\cite{MINDLIN}.
This model is free of arbitrary parameters in the sense that all input 
variables are material properties amenable to experimental verification.
The model involves four key elements: 
1) particle elasticity, 2) energy
loss through internal friction, 3) attraction on the contact surface and 
4) energy loss through the action of frictional forces.

\subsection{Particle Elasticity}

As demonstrated by H. Hertz \cite{HERTZ}, the deformation of elastic 
particles of finite extent leads to a nonlinear
dependence between the compressive force and the compression length.

\smallskip
\begin{equation}
         {\bf F}_{ij}^{\rm elastic} = {4\over 3}\,
	   	{E_i\, E_j\over E_i\, (1-\nu_j^2)+ E_j\, (1-\nu_i^2)}
                        \,\sqrt{R_iR_j\over R_i+R_j} \, h_{ij}^{3/2} \,
			{\bf n}_{ij}^{\perp}, 
   \label{eq:elastic}
\end{equation}
\ni where, 
\begin{equation}
         h_{ij} = R_i+R_j - \mid {\bf X}_i -{\bf X}_j \mid, 
\end{equation}
\ni and, 
\begin{equation}
         {\bf n}_{ij} = \frac{{\bf X}_i -{\bf X}_j}
					    {\mid {\bf X}_i -{\bf X}_j \mid}.
\end{equation}
\smallskip

\ni 
Here, $R_i$ symbolizes the radius of the $i$-th particle.
$E_i$ represents the elastic modulus and $\nu_i$ the Poisson ratio.
${\bf X}_i$ is the position vector of the $i$-th particle and 
${\bf n^{\perp}_{ij}}$ indicates a unit
vector pointing from grain $j$ to grain $i$ perpendicular to the contact
surface.
(Because all the forces described here are contact forces, 
they are nonzero only if $h_{ij}>0$.)

\subsection{Internal Friction}

Energy dissipation due to the viscous nature of the solid particles was
first studied by Kuwabara and Kono \cite{KK} more than a century after 
Hertz's work.
They obtained the following expression for the non-conservative viscous force 
acting during a collision:

\begin{equation}
         {\bf F}_{ij}^{\rm viscos} = -2
	   	{B_i\, B_j\over B_i\, (1-\sigma_j^2)+ B_j\, (1-\sigma_i^2)}
                        \,\sqrt{R_iR_j\over R_i+R_j} \, h_{ij}^{1/2} \,
			{\bf v}_{ij}^{\perp}, 
\end{equation}
\ni where, 
\begin{equation}
         B_i=\frac{9\xi_i\eta_i}{3\xi_i+\eta_i}
\end{equation}
\ni and, 
\begin{equation}
         \sigma_i=\frac{3\xi_i-3\eta_i}{2(3\xi_i+\eta_i)}.
\end{equation}

\ni $\xi_i$ and $\eta_i$ are the coefficients of viscosity associated with 
volume deformation and shear. ${\bf v_{ij}}^{\perp}$ is the relative 
velocity of the colliding particles normal to the contact surface.

\subsection{Surface Attraction}

When two surfaces are brought into contact an attractive
force due to the attractive part of the inter-molecular interaction arises.
The case of spherical particles composed of molecules interacting via a 
Lennard-Jones potential was first studied by Hamaker \cite{HAM} for 
non-elastic grains. It was extended to the case of elastic grains by Dahneke
\cite{DAHN}. The form
used here was introduced by Johnson et al. \cite{MESO} for a general molecular 
interaction characterized by
the surface energy, $W_{ij}$, of the contacting materials.

\smallskip
\begin{equation}
         {\bf F}_{ij}^{\rm meso} = -\sqrt{ {4\over 3}\, 
{8\pi\, W_{ij}\,
                E_i\, E_j\over E_i\, (1-\nu_j^2)+ E_j\, (1-\nu_i^2)} }
                        \,\bigl({R_iR_j\over R_i+R_j}\bigr)^{1/4} 
                  \, h_{ij}^{3/4} \,
                        {\bf n}_{ij}^{\perp}
\end{equation}
\smallskip

\ni Note the small power, $1/4$, to which the 
reduced radius is raised. Since the weight of a spherical grain is equal
to $4/3\pi \rho g R^3$, there will be a grain
size for which the weight of a particle is equal in magnitude to 
this attractive force. Particles smaller than this critical size 
experience this interaction as an adhesive. Typically, this
critical grain size is on the order of 1 mm \cite{KOH1}. For grain
sizes much larger than about 1 cm, this force can be neglected.

\subsection{External Friction Forces}

The frictional forces which develop under conditions of slip parallel to, or
rotation about the normal to the contact surface were first studied by
Mindlin \cite{MINDLIN}. Mindlin's original expression is computationally 
expensive, therefore
a  simplified expression is used which amounts to assuming that no partial 
slipping of the contact surfaces occurs.

\smallskip
\begin{eqnarray}
         {\bf F}_{ij}^{\rm shear} = \min\Biggl( {16\over 3}\,
	   	{G_i\, G_j \, \delta s \over G_i\, (2-\nu_j)+ G_j\, (2-\nu_i)}
                        \,\sqrt{R_iR_j\,\over R_i+R_j} \,h_{ij}^{1/2}\, , 
\nonumber \\
\mu \mid {\bf F}^{\perp}\mid \Biggr)  
           \frac{-{\bf v}_{ij}^{\parallel}}{\mid {\bf v}_{ij}^{\parallel}\mid}
		\label{eq:shear}
\end{eqnarray}
\smallskip

\ni $G_i$ is the shear modulus of the material and $\mu$ is the static friction
coefficient. $\delta s$ is the integrated slip in the shearing direction since 
the particles first came into contact. $\delta s$ is allowed to increase 
until the shearing force exceeds
the limit imposed by the static friction. At that point the contact slips 
and $\delta s$ is reset to zero.

Walton and Braun \cite{MINDLIN} were the first who attempted to incorporate 
Mindlin's 
friction theory into their simulations using what they called
the ``incrementally slipping friction model''. Their model
is computationally more expensive than eq. \ref{eq:shear}, however, the
qualitative results appear to be the same.

When there exist a relative rotation about the
axis perpendicular to the contact surface, then the frictional forces
will induce a moment to counteract this rotation. In Mindlin's theory
this induced moment is described by the following equation when partial
slipping is ignored:

\smallskip
\begin{eqnarray}
          M_{ij}^{\rm couple} = \min\Biggl({8\over 3}\,
	   	{G_i\, G_j \, \delta \alpha \over G_i\, (2-\nu_j)+ G_j\, (2-\nu_i)}
                        \,\left[ \frac{R_iR_j}{R_i+R_j}\right]^{3/2} 
				\,h_{ij}^{3/2}\, , 
\nonumber \\
\, \frac{3\pi}{16}\sqrt{R_iR_j\,\over R_i+R_j} \,h_{ij}^{1/2}
\mu \mid {\bf F}^{\perp}\mid \Biggr) 
           [-{\rm sign}(\omega^{\perp})]\label{eq:couple}
\end{eqnarray}
\smallskip

\ni $\omega^{\perp}$ is the velocity about the axis perpendicular to the
contact surface. $\delta \alpha$ is the integrated angular slip. It plays
the same role as $\delta s$ does for slip along the shear direction.

\subsection{The Complete Model}

In addition to the couple describe by eq. \ref{eq:couple} there will be 
torques induced by the action of the shear forces, given in eq. \ref{eq:shear}.
${\bf F}_{ij}^{\rm elastic}$, ${\bf F}_{ij}^{\rm viscos}$ and ${\bf
F}_{ij}^{\rm meso}$ do not produce any torques because they act radially.

In total, the model described here depends upon eight material 
parameters. (In addition to the seven identified above, the density of the 
material making up the particles is needed in order to calculate the masses
used for integrating Newton's equations of motion.)
Although all of these parameters are in principle measurable, their
determination may, in some cases, be problematic. Measuring the internal
viscosities, for example, is a difficult task. Fortunately, it is not
necessary to know these parameters to high accuracy in order to obtain good
results from the simulations.

\subsection{A Note On the Algorithm}

The computational procedure used here is a generalization of the molecular
dynamics method and consists of calculating the
positions and velocities of each particle at every time step \cite{AT}. As a
prerequisite the forces acting between all pairs of particles must
be calculated using the procedure described in the proceeding section. 
The resulting forces are then employed to integrate
the Hamiltonian form of Newton's equations of motion. 

Modern workstations, and modern parallel computers, are moving
to cache based systems in order to overcome the ever increasing gap between 
processor speed and memory speed. Using such systems in an efficient
manner poses a different set of problems compared to those faced
when using vector machines. In particular data locality and cache reuse are 
the major concerns on cache based systems.

The algorithm used in the present studies attempts to addresses these 
issues.  A complete description of the program will be given elsewhere, 
here it suffices to mention the program's efficiency in terms of its 
execution speed on various platforms.

\begin{table}[htbp]
\begin{center}
 
\begin{tabular}{c|c}\hline\hline \small
    processor   &Speed (Kups) \\\hline
Sparc-10 &   24.1\\
Sparc-20 &   37.5\\
SG R4400 &   41.3\\
IBM-Power2 & 45.2\\
\hline\hline
\end{tabular}

\caption{\small This table gives the program speed (in thousands of particle 
updates
per second) on various processors. For these
test, a dense system consisting of 9261 particles in a 3-D cube with periodic 
boundary conditions on all sides was used.}

\end{center}
\end{table}

By way of comparison, one of the fastest 2-D programs mentioned in the 
literature runs at 28.2 Kups (thousands of particle updates per second) 
on a Sun Sparc-10 \cite{FORM}. That 2-D program 
was ported 
to a Sparc-10 after having been optimized for a vector machine. By comparison,
the present 3-D program was optimized for data locality and cache reuse and
the speed is given in table 1.

\section{Description of the Experiment}

\begin{figure}[htbp]
   \centerline{
               \epsfysize=8cm
                \epsfbox{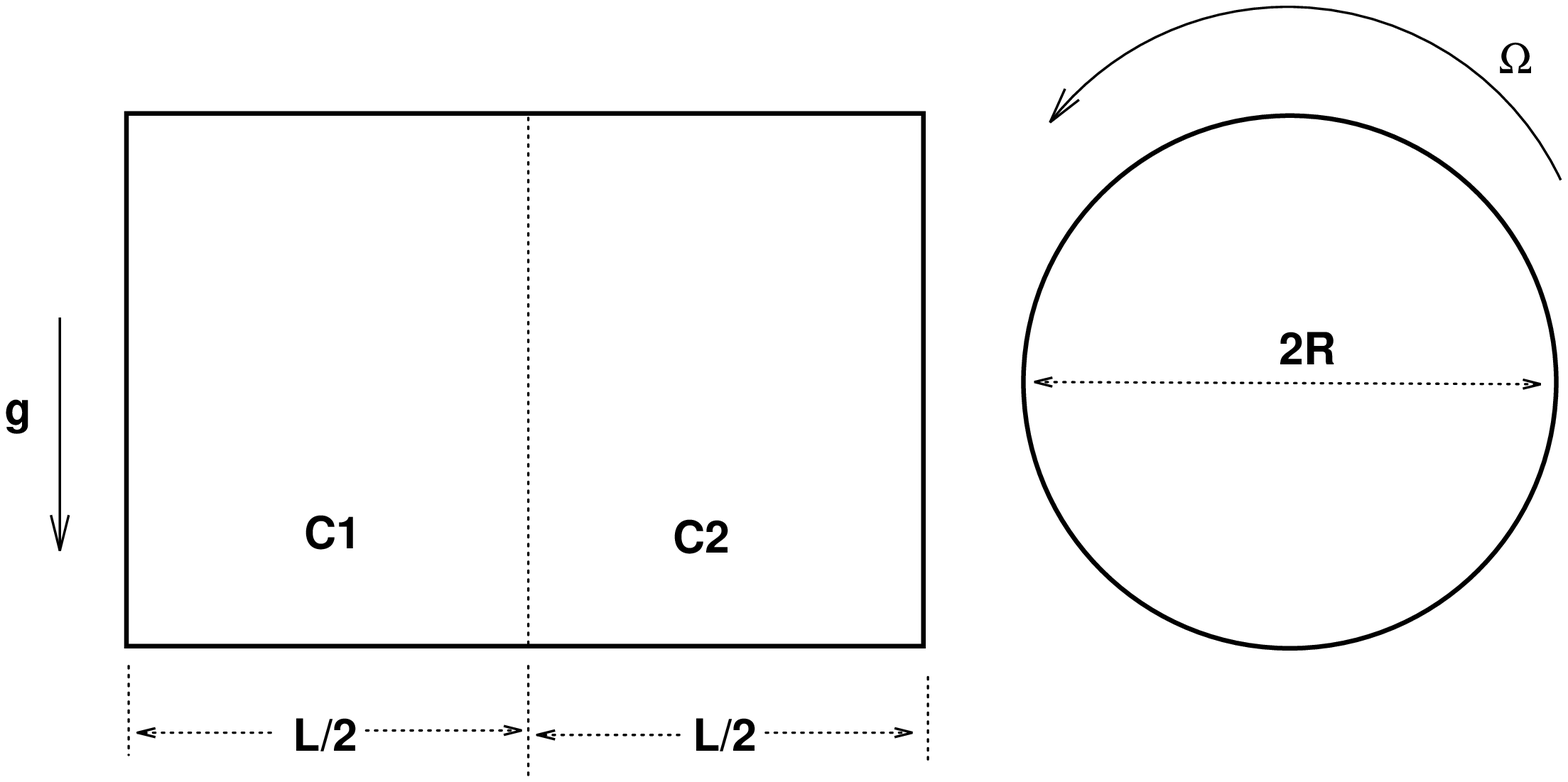}
              }
    \caption{Schematic illustrating the experimental set-up. Initially the
             two material components, C1 and C2 are well separated.
              \label{fig:initposit}
            }
\end{figure}

The diffusional processes inside a rotating drum can be studied using an
arrangement first suggested by Hogg et al. \cite{HOGG}: A drum oriented 
parallel to the
ground is partially filled with two types of
distinguishable particles. One type in the left half and one type
in the right half as shown in fig. \ref{fig:initposit}

As the drum rotates about its axis at a constant speed, momentum is
imparted to the particles in the plane perpendicular to the axis of
rotation.  Motion along the longitudinal 
direction occurs only through random momentum changes induced by collisions
between particles. Hence, motion parallel to the axis of rotation is 
diffusive. 

The diffusion equation for the concentration of each component can be written
as:

\begin{equation}
     \frac{\partial C_1(z,\N)}{\partial \N} = 
           D\frac{\partial^2 C_1(z,\N)}{\partial z^2},
      \label{eq:diffeq}
\end{equation}
\smallskip

\ni where the natural definition of time for this system has been used, namely:
the number of revolutions, $\N$. $C_1(z,\N)$ is the 
concentration of component 1 along the rotation axis.

Eq. \ref{eq:diffeq} can be solved subject to particle conservation and to 
the initial condition discussed above, yielding \cite{HOGG}:

\begin{eqnarray}
C_1(z,\N) = \frac{1}{2} + \frac{2}{\pi}\sum_{n=1}^{\infty}\frac{1}{2n-1}
 \exp\left[-\frac{(2n-1)^2\pi^2D\N}{L^2}\right]
\sin\left[\frac{(2n-1)\pi z}{L}\right] \nonumber \\
\label{eq:diffsol}
\end{eqnarray}

\ni A similar result is obtained for the second component. In the 
computer simulations the average location of all the particles 
belonging to a particular component can be measured with a higher
accuracy than the particle concentration as a function of position. 
From eq. \ref{eq:diffsol}, the average position is:

\begin{eqnarray}
    <z>_1 \, =  \frac{8L}{\pi^3}\sum_{n=1}^{\infty}\frac{(-1)^{n+1}}{(2n-1)^3}
               \exp\left[-\frac{(2n-1)^2\pi^2D\N}{L^2}\right]
\label{eq:avgfull}
 \\
           \approx 
\frac{8L}{\pi^3}\exp\left[-\frac{\pi^2D\N}{L^2}\right] 
\qquad \qquad \qquad \qquad \qquad
\label{eq:avgpos}
\end{eqnarray}
\smallskip

\ni Eq. \ref{eq:avgpos} is a good approximation to eq. \ref{eq:avgfull}.
It can easily be checked that all the higher order terms together contribute 
only about 3\% to the average at $\N=0$. For larger $\N$, the higher order
terms become completely negligible. Hence, by measuring $<z>$ as a 
function of $\N$ one can via eq. \ref{eq:avgpos} obtain the diffusion
constant, $D$.

The computer experiments make use of soft spheres, i.e., particles with 
an elastic modulus, $E \sim 10^6\, {\rm N/m}^2$. It is 
possible to actually manufacture and use particles with such low
elastic moduli in laboratory experiments \cite{MESO}. Indeed they are 
valuable for 
studying inter-particle interactions. The purpose of using them in the
present simulations is to save computer time, since the integration time 
step must be smaller than the collision time which varies like 
$t_c\sim\sqrt{m/E}$ \cite{HERTZ} for particles of mass $m$. Using a normal 
value of 
$E \approx 10^{11}$ would have increased the cpu time 300-fold. 

Table \ref{tab:material} list the material parameters and table
\ref{tab:tribology} list the tribological parameters used in the present
simulations. As can be seen the softness of the particles is reflected not
only in the elastic modulus, but also in the internal viscosities.

\begin{table}[htbp]
\begin{center}
 
\begin{tabular}{c|c|c|c}\hline\hline \small
Parameter  &Material-1 & Material-2 & Material-3\\\hline
$E$        & $1.0\times 10^6\  {\rm N/m^2}$ & $1.0\times 10^6\  {\rm N/m^2}$ 
& $1.0\times 10^6\  {\rm N/m^2}$ \\
$G$        & $0.3\times 10^6\  {\rm N/m^2}$ & $0.3\times 10^6\  {\rm N/m^2}$ 
& $0.3\times 10^6\  {\rm N/m^2}$ \\
$\nu$      & 0.25                         & 0.25                         
&0.25\\
$\xi$      & 5000 poises                   & 5000 poises                   
& 5000 poises \\
$\eta$     & 5000 Poises                   & 5000 poises                   
& 5000 Poises \\
$\rho$     & 1000 ${\rm kg/m^3}$          & 1000 ${\rm kg/m^3}$          
& 1000 ${\rm kg/m^3}$  \\
$R$        & 0.0036 m                     & 0.0024 m                     
& 0.0030 m    \\
\hline\hline
\end{tabular}
 
\caption{\small 
\label{tab:material}
Material parameters used for the soft spheres in the present simulations. 
(See the text for an explanation of the symbols.)
}
 
\end{center}
\end{table}

\begin{table}[htbp]
\begin{center}
 
\begin{tabular}{c|c|c|c|c}\hline\hline \small
Parameter  & 1 - 1  & 1 - 2 & 2-2 &1-3, 2-3, 3-3 \\\hline
$\mu$      & 0.1    & 0.2 & 0.5 & 0.2\\
$W$        & 0.2 ${\rm J/m^2}$ & 0.2 ${\rm J/m^2}$ & 0.2 ${\rm J/m^2}$
&0.2 ${\rm J/m^2}$ \\
\hline\hline
\end{tabular}
 
\caption{\small 
\label{tab:tribology}
Tribological parameters used for the soft spheres in the present simulations. 
$i-j$ indicates the value the parameter takes when a particle of material type
$i$ is in contact with a particle of material type $j$.
(See the text for an explanation of the symbols.)
}
 
\end{center}
\end{table}

The walls of the mixer are also composed of particles in order to reduce the 
complexity of the force calculation. In the present simulations material-3
was used exclusively for the drum walls and materials 1\& 2 where  used for
the particles inside the drum.

Note, the difference in $\mu$ for the three materials.
Previous experimental studies of rotating drums indicate that materials 
with significantly different angles of repose, may segregate rather than
mix during the course of the experiment \cite{MATSEG}. In the simulations, 
differences in
the angle of repose are achieved by varying either $\mu$, the coefficient
of static friction or by varying $W$, the surface energy. Since common
experiments use mixtures of sand and glass beads, varying $\mu$, while
keeping $W$ constant should yield results which are more readily comparable 
with experiments.

The cylindrical drum used in the present simulations was 12 cm long and 8 cm 
in diameter, giving an aspect ratio of 1.5. At the beginning of the simulation,
equal masses of particles from materials 1 or 2 were placed in the drum
as described above. The drum was then rotated at a constant angular speed.
A typical simulation involved about 1000 particles inside the drum and 1000
particles making up the drum walls.

\section{Results and Discussion}

\begin{figure}[htbp]
   \centerline{
               \epsfysize=12cm
                \epsfbox{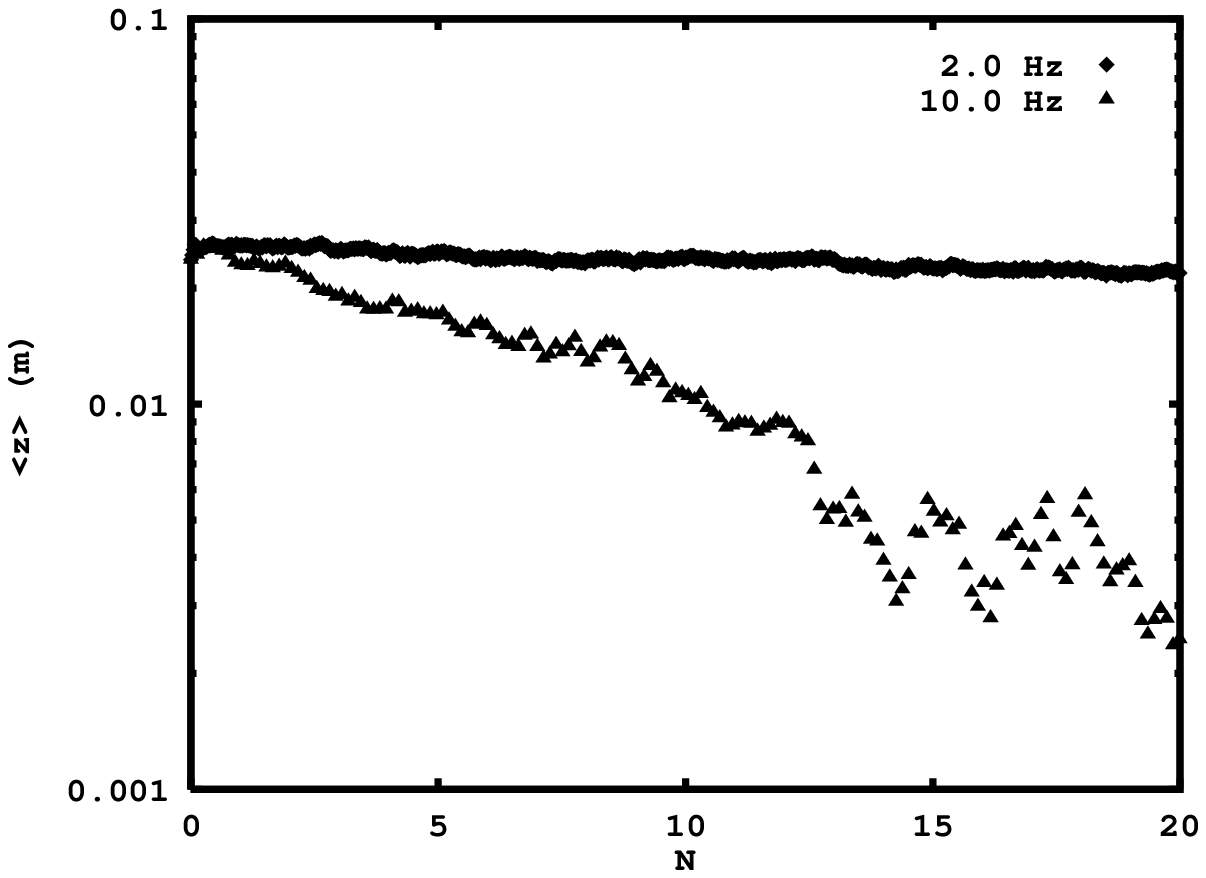}
              }
    \caption{Self-diffusion: Average position as a function of the number
             of revolutions $\N$ for two different rotation speeds.
              \label{fig:avgz11}
            }
\end{figure}

As a first step the self-diffusion for particles composed of material 1
is studied. (This is done by giving each particle a tag corresponding to
that half of the drum in which it started.) The average position in the
longitudinal direction (denoted by: $z$) is plotted in fig. \ref{fig:avgz11}
as a function of the number of drum revolutions for two different 
rotation speeds. As to be expected,
the particles diffused faster at higher rotation speeds. 

A further understanding of the diffusion process 
can be obtained by studying animated videos of the rotating drum.
The videos show quite clearly that diffusion is initiated at the
free surface of the granulate, diffusion in the bulk plays a subordinate
role for the case of a single species. Basically, particles brought up to
the free surface via the action of the drum's rotation execute a random 
walk parallel to the axis of rotation as they roll along the surface. 

\begin{figure}[htbp]
   \centerline{
               \epsfysize=12cm
                \epsfbox{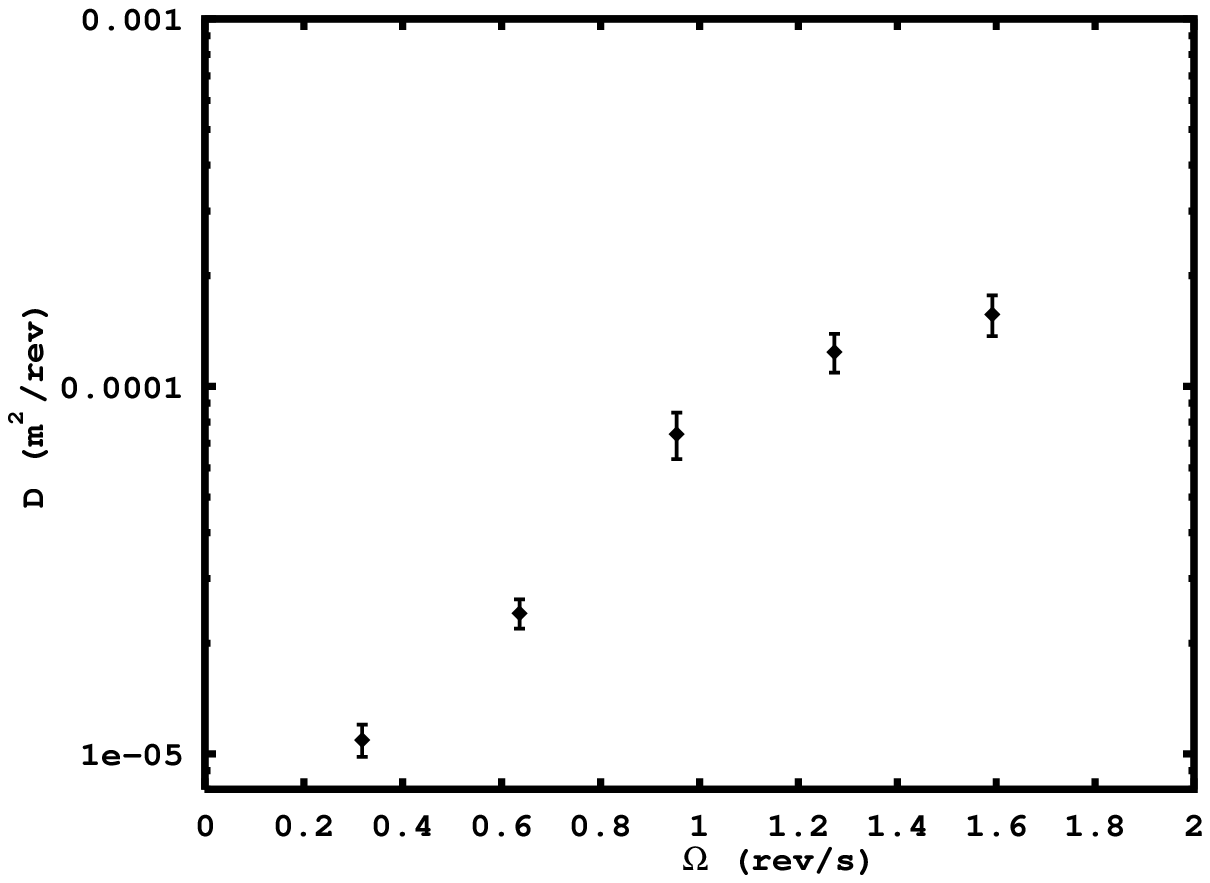}
              }
    \caption{Diffusion coefficients for self-diffusion as a function of the
             number of revolutions.
              \label{fig:selfdcon}
            }
\end{figure}

From fig. \ref{fig:avgz11} and similar plots the diffusion constants can 
be extracted. A plot of the self-diffusion
coefficient as a function of rotation speed is shown in fig. \ref{fig:selfdcon}.
This data agrees qualitatively with the experimental results of Rao et al.
\cite{RAO}.
Quantitative agreement is not expect because Rao et al. are using particles
with different material properties.

\begin{figure}[htbp]
   \centerline{
               \epsfysize=12cm
                \epsfbox{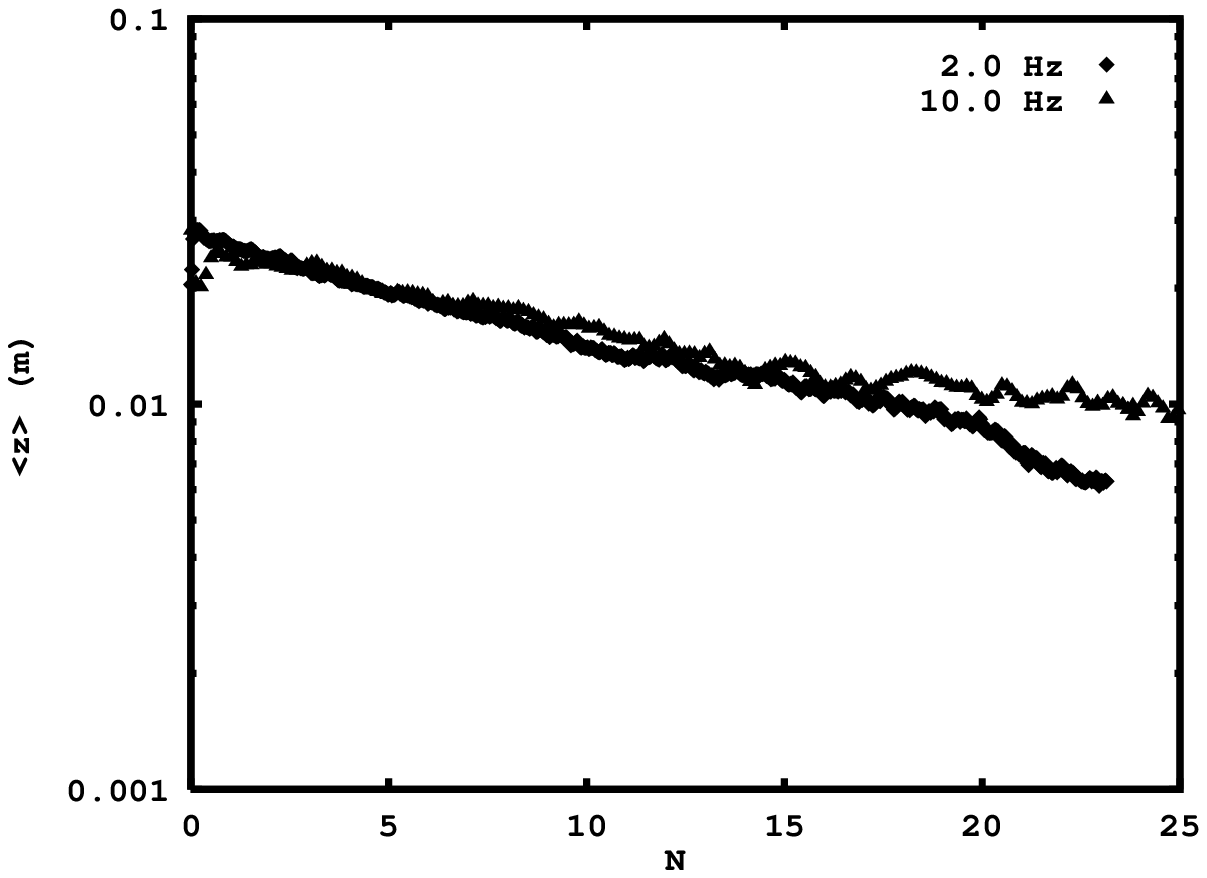}
              }
    \caption{2-Component diffusion: Average position as a function of the 
            number of revolutions $\N$ for two different rotation speeds.
              \label{fig:avgz12}
            }
\end{figure}

\begin{figure}[htbp]
   \centerline{
               \epsfysize=12cm
                \epsfbox{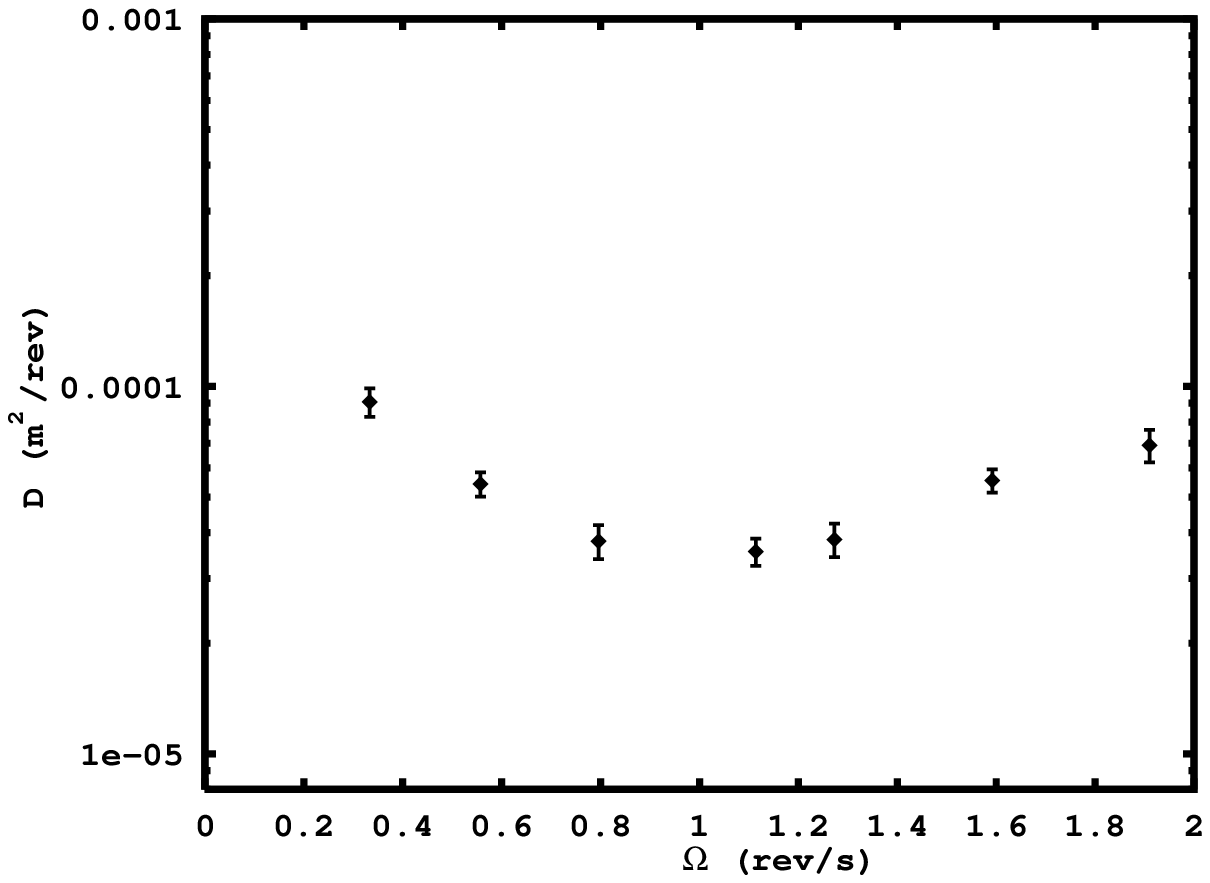}
              }
    \caption{Diffusion coefficients for 2-component diffusion as a function 
             of the number of revolutions.
              \label{fig:dcon}
            }
\end{figure}

The case of two diffusing species has not, to our knowledge, been studied
experimentally. Fig. \ref{fig:avgz12} depicts the average position of the
particles composed of material 1 as they diffuse through particles composed
of material 2. Note the qualitative difference between this figure and
fig. \ref{fig:avgz11}. For the two-species situation, the diffusion rate is
nearly the same for the two rotation speeds and in fact it is slightly smaller
at the higher rotation rate. This counterintuitive phenomenon exists over a
range of rotation speeds as illustrated in \ref{fig:dcon}.

Again, animated videos shed some light on the processes taking place. Since
the particles of material 2 have a larger coefficient of static friction than
those of material 1, their angle of repose is larger, which means that 
they will tend to roll downhill onto the particles of material 1. 
The particles of material 2 are also smaller than those of material 1, hence
they are able to diffuse through material 1 not only at the surfaces, but
also in the bulk. (As layers of species 1  move past each other they open
up voids through which the smaller particles can move.) For the two-species 
case, diffusion in the bulk plays a larger role than for the single-species 
case. 

\begin{figure}[htbp]
   \centerline{
               \epsfysize=12cm
                \epsfbox{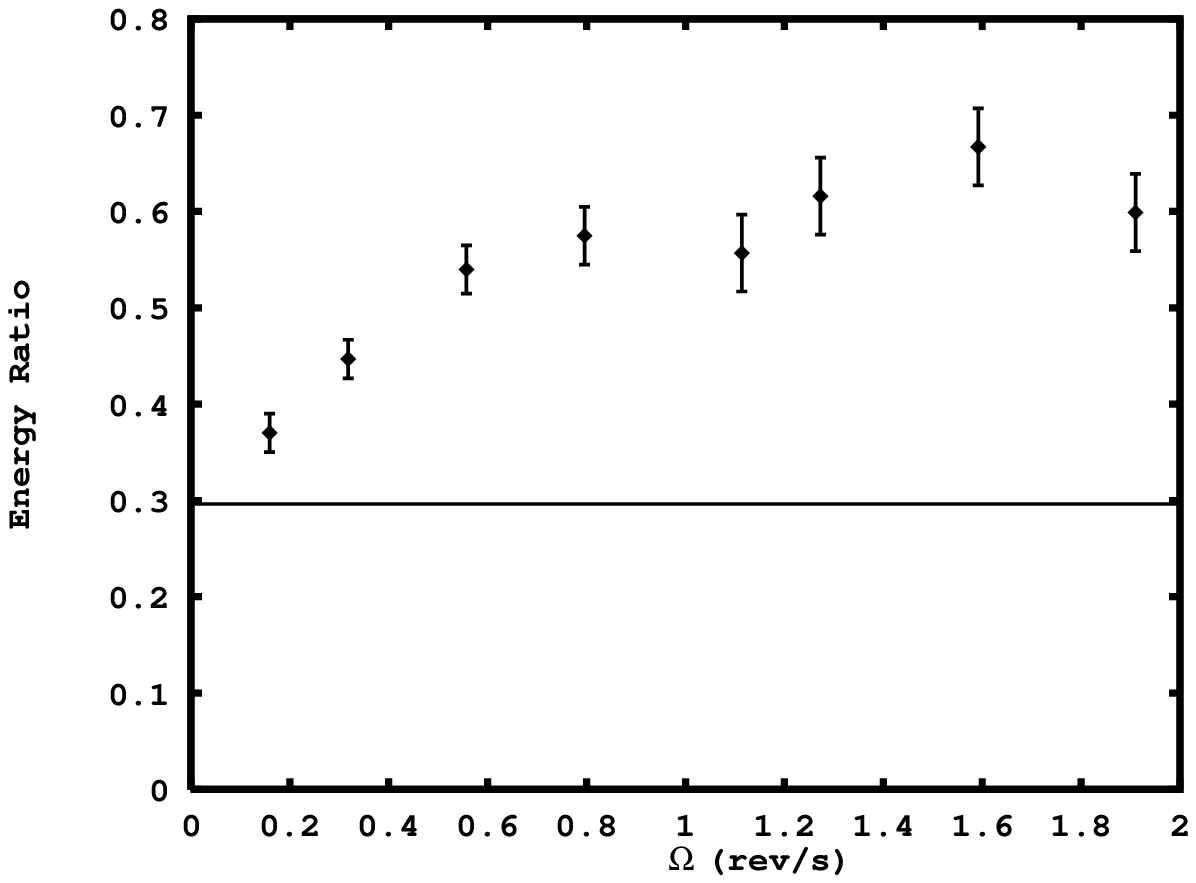}
              }
    \caption{The ratio of the average kinetic energy for particles composed
             of material 2 to that for particles composed of material 1.
              \label{fig:ergratio}
            }
\end{figure}

Another indicator of the enhanced mobility of the smaller particles is
given by the ratio of the total kinetic energy of the small particles to that
of the larger particles. Fig. \ref{fig:ergratio} shows this ratio as a 
function of the rotation speed. The solid line indicates the expected ratio if
the average velocity of particles belonging to 
both species were equal. As can be
seen, the measured ratios lie above this line, thus the smaller particles
are moving, on average, with higher speeds than the larger particles. In other
words, the diffusion in this two-component system is being driven by the 
smaller particles.

At larger rotation speeds particles roll down the surface faster leaving
less time for motion parallel to the rotation axis. Hence, the diffusion 
processes are slowed because their are fewer particles from species 2
rolling onto those of species 1. Eventually, as the rotation speed increases,
the entire free surface will fluidize allowing for increased mobility
in the longitudinal direction and the diffusion coefficient increases.

If the rotation speed is increased still further, one eventually enters the
centrifugal regime where the particles are held to the inner surface of the
drum via centrifugal forces. In this regime all diffusion must come to a
halt. This implies that there must be a maximum diffusion coefficient at some
rotation speed larger than those which could be studied here.

Likewise, as the rotation speed decreases towards zero the mobility of the 
small particles must also decrease until the diffusive motion stops,
consequently, there must also be a maximum diffusion coefficient at some
rotation speed much smaller than those which could be studied here.

\section{Summary and Conclusions}

A computational model for the collision of two visco-elastic
spheres which is independent of arbitrary parameters has been presented. 
Using this model the diffusion of particles in a three dimensional rotating 
drum has been studied. The major result is that the diffusion coefficient
for a two-component system has a much richer structure than anticipated by
studying a single-component system. This is primarily due to the larger
mobility of the smaller particles compared to that of the larger particles.

One phenomena in rotating drums which has been given a good deal of 
attention in the literature is the spontaneous segregation of two 
material types \cite{MATSEG}. In a typical experiment, one component is 
smaller and
rougher (and thus has a larger angle of repose) than the other. After
about 5 minutes of rotation, the two materials spontaneously segregate
into a series of bands parallel to the rotation axis. The composition
of the bands alternates between high concentrations of component 1 and
high concentrations of component 2. In most cases there is an odd number
of bands, with the smaller, rougher particles located at the ends of the
drums.

The mechanisms which create this banded structure are fairly well understood. 
A statistical
fluctuation in the concentration of the smaller, rougher component yields a
local increase in the angle or repose. The larger particles will tend to
roll away from such a locality, thus depleting their numbers in the region
of the fluctuation which drives the angle of repose to even higher values.
Now, it is not completely understood why the bands should be stable, since 
as shown here, particle diffusion will tend to destroy the
banded pattern.

This phenomena occurs on a time scale which is too large to be simulated
with present resources, however, the above results can be applied to this 
problem. The present work has shown that diffusion in two-component systems
is driven by the mobility of the smaller particles. Therefore it is possible
to understand the stability of band patterns as follows. Let a
band of large particles exist between two bands of smaller particles. If
the rate at which the small particles diffuse through the larger particles is
large enough, then the small particles from one band can diffuse through
the band of large particles and replenish the band of small particles on
the other side. Now if the current of small particles is conserved, then
diffusion of the large particles into the region occupied by the small
particles will be depressed due to geometrical factors.

This idea implies that only systems with an odd number of bands are stable,
as is indeed reported in the literature. (There has been only one reported
experiment yielding an even number of bands \cite{MATSEG}, however they 
did not use a
simple cylindrical drum, rather they used a more complicated drum shape
which enhanced the angle of repose of the larger particles.) Furthermore 
it explains why
early experiments claimed that the segregation started with the smaller
particles at the ends of the drum. Indeed any other arrangement would 
necessarily be unstable.

It may be possible to test these and other ideas on spontaneous material
segregation in the near future through the use of high performance
computer systems.

\newpage

\section{Figures}

\bigskip
\bigskip

\bigskip

\bigskip

\bigskip

\bigskip

\bigskip

\bigskip

\bigskip

\end{document}